\documentclass[12pt]{article}
\pdfoutput=1
\usepackage{jcappub}

\usepackage{url}

\newcommand{\iu}{\mathrm{i}}
\newcommand{\me}{\mathrm{e}}
\newcommand{\dd}[1]{~\mathrm{d}#1~}
\newcommand{\unit}[1]{\,\mathrm{#1}}
\newcommand{\factor}[3]{\left(\frac{#1}{#2\unit{#3}}\right)}

\makeatletter
\def\dv{\@ifstar\@dv\@@dv}
\newcommand{\@@dv}[3][]{\frac{\mathrm{d}^{#1}#2}{\mathrm{d}#3^{#1}}}
\newcommand{\@dv}[3][]{\mathrm{d}^{#1}#2/\mathrm{d}#3^{#1}}
\makeatother

\renewcommand{\vec}[1]{\boldsymbol{#1}}

\def\epsilon{\varepsilon}
\def\theta{\vartheta}
\def\lsim{\raise0.3ex\hbox{$\;<$\kern-0.75em\raise-1.1ex\hbox{$\sim\;$}}}
\def\gsim{\raise0.3ex\hbox{$\;>$\kern-0.75em\raise-1.1ex\hbox{$\sim\;$}}}

\def\PKS{\mbox{PKS\:2155-304}}

\newcommand{\apj}{{Astrophys.\ J. }}
\newcommand{\apjl}{{Astrophys.\ J.\ Lett. }}

\title{Detecting ALP wiggles at TeV energies}

\author{M. Kachelrie{\ss} and}
\author{J. Tjemsland}

\affiliation{Institutt for fysikk, NTNU, Trondheim, Norway}


\abstract{
  Axions and axion-like-particles (ALPs) are characterised by their two-photon
  coupling, which entails so-called photon-ALP oscillations as photons
  propagate through a magnetic field. These oscillations lead
  to distinctive signatures in the energy spectrum of high-energy photons
  from astrophysical sources, allowing one to probe the existence of ALPs.
  In particular, photon-ALP oscillations will induce energy dependent
  oscillatory features, or ``ALP wiggles'', in the photon spectra. We propose
  to use the discrete power spectrum to search for ALP wiggles and
  present a model-independent statistical test.
  By using \PKS\ as an example, we show that the method has the
  potential to significantly improve the experimental sensitivities for ALP
  wiggles, and that the ALP wiggles
  may be detected using the Cherenkov Telescope Array (CTA)
  for optimistic values of the photon-ALP coupling constant and the magnetic
  field.
  Moreover, we discuss how these sensitivities depend on the modelling of the
  magnetic field. We find
  that the use of realistic magnetic field models, due to their larger cosmic 
  variance, substantially enhances detection prospects compared to the use of
  simplified models.
}

\begin{document}

\maketitle

\section{Introduction}

Axions are well motivated beyond the standard
model particles that can explain a variety of unsolved problems in physics,
such as the strong CP
problem~\cite{Peccei:1977hh,Peccei:1977ur} and the nature of dark
matter~\cite{Preskill:1982cy,Abbott:1982af,Dine:1982ah}.
These particles are mainly characterised by their two-photon coupling
$g_{a\gamma}$ from the interaction term
$\mathcal{L}=\frac{1}{4}g_{a\gamma}aF_{\mu\nu}\tilde{F}^{\mu\nu}
=g_{a\gamma}a\vec{E}\cdot\vec{B}$, and by their small mass $m_a$
obtained through pion mixing~\cite{Weinberg:1977ma,Wilczek:1977pj}.
The relationship between $g_{a\gamma}$ and $m_a$ is thus fixed as
$g_{a\gamma}\unit{GeV}\sim 10^{-16}m_a/\unit{\mu eV}$ up to a
$\mathcal{O}(1)$ factor~\cite{GrillidiCortona:2015jxo,DiLuzio:2020wdo}.
A more general class of light pseudoscalar particles which share the same
two-photon coupling as the axion but have an arbitrary mass $m_a$, is known as
axion-like particles (ALPs).
Although ALPs do not solve the strong CP problem, they are
nevertheless interesting as they, e.g., arise naturally in string theories and
other extensions of the standard 
model~\cite{Arvanitaki:2009fg,Cicoli:2012sz,Alexander:2023wgk}.

The majority of ALP searches are based on photon-ALP mixing in a magnetic field
(see Refs.~\cite{Graham:2015ouw,Irastorza:2018dyq} for two recent reviews):
Due to the characteristic two-photon--ALP vertex, a photon/ALP may interact
with a virtual photon provided by the magnetic field and convert into an
ALP/photon.
Currently, the most solid and extensive exclusions at sub-eV masses, 
$g_{a\gamma}<6.6\times 10^{-11}\unit{GeV^{-1}}$, are set by the CAST helioscope
($m_a\lesssim \unit{eV}$)~\cite{CAST:2017uph} by attempting to convert solar
ALPs into photons on Earth. A comparable limit is found for
$m_a\lesssim \unit{keV}$ by studying the lifetime of horizontal branch
stars~\cite{Ayala:2014pea,Dolan:2022kul}.
The planned ``shining light through a wall'' experiment
ALPS-II~\cite{Bahre:2013ywa} and the solar helioscope IAXO~\cite{IAXO:2019mpb}
are expected to improve upon these limits immensely.
Significantly stronger limits around $m_{a\gamma}\sim 10^{-6}\unit{eV}$
are obtained for ALP dark matter in haloscope experiments, such as
ADMX~\cite{ADMX:2019uok} and the upcoming ABRACADABRA
experiment~\cite{Ouellet:2018beu}, or by conversion near neutron
stars~\cite{Foster:2022fxn,Battye:2023oac}.

The strongest limits
($g_{a\gamma}\lesssim 10^{-11}$--$10^{-13}\unit{GeV^{-1}}$)
at low masses ($m_a\lesssim 10^{-6}\unit{eV}$) are set by
observations of astrophysical
photon sources using the signatures that photon-ALP
oscillations will imprint on photon spectra:
First, photon-ALP oscillations will induce ``irregularities''.
The non-detection of such spectral irregularities has been used to constrain the
parameter space using, e.g., gamma-ray observations by HESS~\cite{HESS:2013udx}
and Fermi-LAT~\cite{Fermi-LAT:2016nkz}, observations of the Galactic diffuse
gamma-rays by Tibet AS$\gamma$ and HAWC~\cite{Eckner:2022rwf}, and using X-ray
observations from
Chandra~\cite{Reynes:2021bpe,Matthews:2022gqi,Reynolds:2019uqt}. 
The Cherenkov Telescope Array (CTA) is expected to improve the limits
from HESS and Fermi-LAT~\cite{CTA:2020hii}.
Second, ALPs that are produced near or in the source can
convert back into photons in, e.g., the Galactic magnetic field, thus inducing 
an additional gamma- or X-ray flux
which has been searched for in SN1987A~\cite{Payez:2014xsa},
Betelgeuse~\cite{Xiao:2020pra}, and super star clusters~\cite{Dessert:2020lil}.
Moreover, ALPs that are sourced near the polar caps in pulsars and resonantly
converted to photons have recently been used to set leading
limits~\cite{Noordhuis:2022ljw}.
Third, photon-ALP oscillations will increase the linear polarisation
of photons, which can be used to set limits using e.g.\ optical photons or
X-rays\footnote{
  The same phenomenon occurs naturally also
  for gamma-rays, but
  the measurement of the polarisation is with current
  and planned detectors not possible~\cite{Galanti:2022yxn}.
}
from magnetic white dwarfs and neutron
stars~\cite{Lai:2006af,Gill:2011yp}.
In Ref.~\cite{Dessert:2022yqq}, it was shown that the measurement of linear
polarisation in white dwarf spectra
excludes $g_{a\gamma}\gtrsim 5.4\times 10^{-12}\unit{GeV}^{-1}$
for $m_a\lesssim 3\times 10^{-7}\unit{eV}$, which is the strongest existing
limit for ALP masses between $\sim 10^{-9}$ and $\sim 10^{-6}\unit{eV}$. At
masses $m_a\lesssim 10^{-11}\unit{eV}$, the best limit is set by the
non-detection of spectral irregularities in X-ray data from
Chandra~\cite{Reynes:2021bpe}. Finally, photon-ALP oscillations will effectively
increase the mean-free path of photons at TeV energies since ALPs travel
practically
without any interactions with the extragalactic background light
(EBL)~\cite{DeAngelis:2007dqd}. This effect has been recently used to set
strong limits with
HAWC~\cite{Jacobsen:2022swa}. Moreover, this effect is important in
combining fit analyses, such as in the recent limit set using
FERMI flat radio quasars~\cite{Davies:2022wvj}.

All the limits discussed in the previous paragraph are, however, strongly
dependent on the treatment of the magnetic
fields~\cite{Kartavtsev:2016doq,Montanino:2017ara,Libanov:2019fzq,%
Carenza:2021alz}.
Therefore, one either needs a reliable description of the magnetic fields, or
knowledge of how uncertainties in the magnetic fields affect the results
(see e.g. the discussions in
Refs.~\cite{Meyer:2014epa,Kachelriess:2021rzc,Matthews:2022gqi}).
This is particularly important for the turbulent component of the magnetic
fields, since oversimplified models are often used to describe these fields. 

In Ref.~\cite{Kachelriess:2021rzc} we introduced the idea of using the discrete
power spectrum to probe photon-ALP oscillations in photon spectra.
In this work, we further discuss and exemplify this concept. In particular,
we introduce a statistical procedure that has the potential to
significantly improve current detection prospects for irregularities
induced by photon-ALP oscillations, which we name ``ALP wiggles''.
The statistical method has two main applications: First, it can be used to
search for ALP wiggles without specifying the EBL distribution
and the magnetic field model. Second, the method is a convenient way to
analyse the effect of various magnetic field models on the expected ALP
wiggles. We find that this method is more robust than a standard $\chi^2$
comparison with data. 

In order to stay as concrete as possible, the examples focus on gamma-rays at
TeV energies, of relevance for the upcoming CTA experiment.
However, the same considerations and discussions
also apply to other energy
ranges, relevant for e.g.~Fermi, HESS and Chandra.
Furthermore, we only include the effect of the
extragalactic magnetic field and fix
$B_0=10^{-9}\unit{G}$ in Sec.~\ref{sec:stat}, while we use
$B_0=5\times 10^{-9}\unit{G}$ in Sec.~\ref{sec:detection}.
Although these are (over-)~optimistic choices for a space-filling,
primordial magnetic field, the magnetic fields in filaments
can easily reach higher values. For example, the turbulent and regular
components of the magnetic fields in galaxies and clusters of galaxies
are expected to be as large as $\mathcal{O}(10^{-6})\unit{G}$.
In addition, we fix $g_{a\gamma}=10^{-11}\unit{GeV^{-1}}$, which 
at $m_a\lesssim 10^{-6}\unit{eV}$ is excluded by a factor of $\sim 2$, and at
$m_a\lesssim 10^{-11}\unit{eV}$ by more than an order of magnitude.
Nevertheless, this choice is appropriate for our purposes:
We advocate for a model independent approach to an ALP searches,
since current limits based on astrophysical photon-ALP oscillations
depend strongly on the modeling of the magnetic fields.
Moreover, it suffice to show that the approach is more sensitive than the
standard $\chi^2$ search for residuals, which have already been used
to set competitive and leading limits~\cite{HESS:2013udx,Fermi-LAT:2016nkz,%
Eckner:2022rwf,Reynes:2021bpe,Matthews:2022gqi,Reynolds:2019uqt}.

\section{Photon-ALP oscillations in astrophysical magnetic fields}

\subsection{Equation of motion}
Physically, one can interpret the photon-ALP oscillation as a mixing
between two mass eigenstates, similar to neutrino oscillations. The mixing
strength and oscillation length depend on the effective mass of the photon which
in turn is determined by the propagation environment (i.e.\ the surrounding
magnetic field, plasma, photon bath, etc.) and photon energy.
A photon and an ALP with energy $E$ propagating in $z$ direction can be
described by the linearised equation of motion~\cite{Raffelt:1987im},
\begin{equation}
  (E+\mathcal{M} - \iu\partial_z)\phi(z) = 0,
  \label{eq:eom}
\end{equation}
where $\phi=(A_\perp, A_\parallel, a)^\mathrm{T}$ is the wave function
describing the two photon polarisation and ALP states.
The mixing matrix can be written as
\begin{equation}
  \mathcal{M} =
  \begin{pmatrix}
    \Delta_\perp & 0 & 0 \\
    0 & \Delta_\parallel & \Delta_{a\parallel} \\
    0 & \Delta_{a\parallel} & \Delta_a
  \end{pmatrix},
\end{equation}
where $\Delta_{\perp/\parallel} = (n_{\perp/\parallel} - 1)E$
($n$ being the refractive index of the photon),
$\Delta_a=-m_a^2/(2E)$ and $\Delta_{a\gamma} = g_{a\gamma}B_\mathrm{T}/2$.
The transverse magnetic field $\vec{B}_\mathrm{T}$ is the component of the
magnetic field perpendicular to the propagation direction, and the
index $\perp$ ($\|$) refers to the direction perpendicular (parallel)
to $\vec{B}_\mathrm{T}$. 

In this work, we consider only photons in the sensitivity range of CTA
($\sim 10^{11}$--$10^{14}\unit{eV}$) and low ALP mass
($m_a\lesssim 10^{-10}\unit{eV}$). Then the dominant contribution
to the photon refractive index is the one from the
EBL~\cite{Kachelriess:2021rzc}, given by~\cite{Dobrynina:2014qba}
\begin{equation}
  \Delta_\mathrm{EBL}\simeq \Delta_\mathrm{CMB}\simeq 0.5\times 10^{-42}E.
\end{equation}
All turbulent magnetic fields lead to similar dependencies on the oscillation
parameters~\cite{Kachelriess:2021rzc}, and the discussions in this
paper can therefore be applied to other energy ranges and magnetic field%
\footnote{
  For example, at X-ray energies and small axion masses, the astrophysical
  photon-ALP oscillations are determined mainly by the plasma density, for which
  $\Delta_\mathrm{pl}\propto E^{-1}$.
}.

\subsection{\tt ELMAG}

We simulate the propagation of photons using
{\tt ELMAG}~\cite{Kachelriess:2011bi,Blytt:2019xad} which is a Monte Carlo
program that simulates electromagnetic cascades of high-energy photons,
electrons and positrons created by their interactions with the EBL. We have
implemented ALPs into
{\tt ELMAG}~\cite{Kachelriess:2021rzc}, thereby allowing for a consistent
treatment of cascading and oscillations. This advantage is however at the
cost of being significantly more computationally demanding than the
alternative Python packages GammaALP~\cite{Meyer:2021pbp} and
ALPro~\cite{Matthews:2022gqi}, which are based on transfer matrices.

Compared to Ref.~\cite{Kachelriess:2021rzc}
we have added the following features {\tt ELMAG}\footnote{
  The code will be made publicly available in a future
  release of {\tt ELMAG}.
}: 
Gaussian turbulent fields with a broken power-law as power
spectrum [see Eq.~\eqref{eq:magnetic_spectrum}] can be modelled,
the magnetic field strength can be distributed as a top-hat function with
a given filling factor, and the
computation time is significantly reduced.

\subsection{Magnetic field models}

High-energy photons will encounter a variety of turbulent magnetic fields
on their path towards Earth, with strengths varying from $B\sim 1$\,G
near jets of AGNe, fields on galactic scales with $\sim \mu$G,
within galaxy clusters ($\sim 0.1$--10\,nG) and finally the intergalactic magnetic
field, see e.g. Refs.~\cite{Batista:2021rgm,Durrer:2013pga} for recent reviews.
The energy in the turbulent magnetic fields in galaxies and galaxy clusters
are believed to be generated at large scales 0.1--10\,kpc through, e.g.,
``mechanical stirring'', large- and small-scale dynamos, and compression.
The energy is in turn
transported to smaller length scales through an energy cascade, leading to a
power-law spectrum of the turbulent magnetic field. It is common to assume
that the magnetic field either has a Kolmogorov ($\gamma=-5/3$) or
Kraichnan ($\gamma=-2/3$) spectrum. At small $k$, a Batchelor spectrum
($\beta=5$) is expected, but other spectral indices have been suggested too.

\begin{figure}
  \centering
  \includegraphics{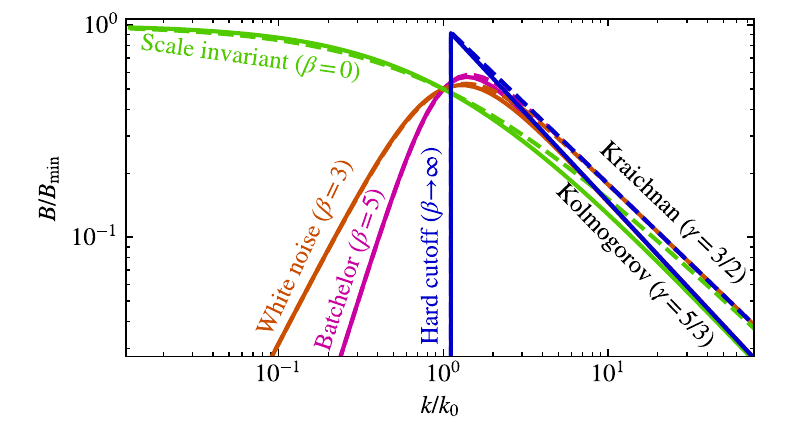}
  \caption{Visualisation of different magnetic field spectra that can be
    modelled in {\tt ELMAG}.}
  \label{fig:magnetic_spectrum}
\end{figure}

In order to take into account the stochastic nature of the turbulence, we
will describe it as a divergence-free Gaussian turbulent field with zero
mean and RMS-value $B_\mathrm{rms}^2=\langle B^2\rangle$. Following the approach
of Refs.~\cite{1994ApJ...430L.137G,1999ApJ...520..204G}, we describe the
magnetic field as a superposition of $n$ left- and right-circular polarised
Fourier modes. The modes will be distributed according to the power-law
spectrum
\begin{equation}
  B_j = B_\mathrm{min} \left(\frac{k_j}{k_0}\right)^{\beta/2} \left[1
    + \left(\frac{k_j}{k_0}\right)^{\gamma+\beta}\right]^{-1/2}
  \label{eq:magnetic_spectrum}
\end{equation}
between $k_\mathrm{min}$ and $k_\mathrm{max}$. The parameter $k_0$ determines
the break in the power law which is visible in the magnetic field spectra
shown in Fig.~\ref{fig:magnetic_spectrum}. In the case of astrophysical
magnetic fields,  $L_0 = 2\pi/k_0$ corresponds to the injection scale.
The field modes extend down to
the dissipation scale $L_\mathrm{min} = 2\pi/k_\mathrm{max}$ which is below
any astrophysical scale of interest. In practise, one cuts off therefore 
the spectrum at a value of $L_\mathrm{min}$ which is much smaller
than the smallest relevant scale of the problem in question\footnote{%
Here, the largest relevant wave number is
$k_\mathrm{osc}=\Delta_\mathrm{CMB}(10^{14}\unit{eV})\sim 8\unit{Mpc^{-1}}$.
}.
We use $k_\mathrm{max}=100k_0$, fix $k_\mathrm{min}$ by the condition
$B(k_{\min})=B(k_{\max})$ and use 33 modes per decade.
The field is normalised such
that $B_\mathrm{rms}^2/2$ coincides with the energy density stored in the
field.  We define the
coherence length $L_\mathrm{c}$ of the turbulent fields as
\begin{equation}
 L_\mathrm{c} = \frac{\pi}{B_\mathrm{rms}^2} \int\frac{{\rm d}k}{k} \vec B^2(k).
\end{equation}
For comparison, we will also consider a simple domain-like field which is
often used in the literature due to its
simplicity~\cite{Reynes:2021bpe,Reynolds:2019uqt,Galanti:2018nvl,%
Galanti:2018myb,Wouters:2012qd,DeAngelis:2011id,Galanti:2018upl,%
Galanti:2018nvl}. 
In this approach, the magnetic field is split into patches with a size
equal to the coherence length $L_\mathrm{c}$. Within each patch, the magnetic
field is homogeneous with a randomly chosen direction. This model is unphysical
and may lead to a bias in the strength of the  ALP signatures
deduced~\cite{Meyer:2014epa,Kachelriess:2021rzc}.

As already mentioned, we will focus on the effect of the intergalactic magnetic
field. From the non-detection of electromagnetic cascades from
blazars, it was concluded that the extragalactic space must be filled with an
turbulent magnetic field with a strength of
$B\gtrsim 10^{-14}$\,G
with a large
filling factor~\cite{HESS:2023zwb,Neronov:2010gir,Dolag:2010ni}, while an upper limit
of $B\lesssim 10^{-9}$\,G is derived from Faraday rotation
measurements~\cite{Pshirkov:2015tua}.
The nature and the production of the extragalactic magnetic field remain
however unknown: A large range of magnetic field
strengths, spectral index at small $k$ (i.e. $\beta$) and coherence lengths are
made possible by the many conceivable
production mechanisms. For example, if produced during inflation, the
initial magnetic spectrum will be scale invariant ($\beta = 0$).
Its coherence length is currently limited by  
hydrodynamical turbulence decay from below ($\sim\unit{kpc}$) and the Hubble
radius from above. Meanwhile, the range of allowed magnetic field
strengths is slowly closing, and it has been argued that the remaining 
parameter space can be completely eliminated by the non-detection of magnetic
halos from misaligned blazars~\cite{Broderick:2018nqf}. As a solution,
Ref.~\cite{Broderick:2018nqf}
proposed that the electromagnetic cascades are quenched by plasma
instabilities, what, if confirmed, would re-open large parts of the parameter
space for intergalactic magnetic fields.

In the remaining of this work, we will focus on the effect of a primordial
intergalactic magnetic field with a field strength $B\sim 10^{-9}\unit{G}$.
This value is chosen to highlight the signatures and the effect of the
statistical method introduced in section~\ref{sec:method}.
Although this value is arguably over-optimistic for primordial fields, similar
field strengths can easily be obtained in filaments between
clusters of galaxies, or in Galactic magnetic fields which we for concreteness
do not include.
Although we consider only TeV photons, all of the discussions and considerations
made in this paper can be applied to other energies and astrophysical
environments~\cite{Kachelriess:2021rzc}, taking into account that the energy
dependence of the refractive index scales as $E^{-1}$ at low energies and as
$E^1$ at high energies.
Due to the many uncertainties, and since the expected signal depends strongly
on the treatment of the magnetic fields, we will in this work advocate for an
experimental approach independent of the modelling of the magnetic fields
and the source spectrum.

\subsection{Parameter space}

Photon-ALP oscillations will  lead to two important signatures on high-energy
photon spectra at $E\sim \unit{TeV}$.
First, they will perturb the photon spectrum by energy dependent
oscillations with $k\sim \Delta_\mathrm{osc}$
[see Eq.~\eqref{eq:homogeneous_solution}], even for a turbulent magnetic
field~\cite{Kachelriess:2021rzc}. Second, the mean free path length of photon
will increase since ALPs will propagate without interacting with the EBL.
In this work, we focus on the former effect. In this subsection, we
will estimate the ALP and magnetic field properties needed to observe ALP
wiggles with CTA, and motivate our focus on intergalactic magnetic fields.
The conditions discussed here can be deduced graphically from
Fig.~3 in Ref.~\cite{Kachelriess:2021rzc}.

For a homogeneous magnetic field, the oscillation probability is given by
\begin{equation}
  P_s(\gamma\to a) =
    \left(\frac{2\Delta_{a\gamma}}{\Delta_\mathrm{osc}}\right)^2
    \sin^2\left(\Delta_\mathrm{osc}s/2\right)
  \label{eq:homogeneous_solution}
\end{equation}
with $\Delta_\mathrm{osc}^2 = (\Delta_\| - \Delta_a)^2 + 4\Delta_{a\|}^2$.
The oscillation length is then defined as
$L_\mathrm{osc} = 2\pi/\Delta_\mathrm{osc}$.
The oscillatory features---which we name ``ALP wiggles''---described by the
solution in Eq.~\eqref{eq:homogeneous_solution} are present also
in turbulent magnetic fields provided that the coherence length 
is on the same order of magnitude or larger than the oscillation length. At TeV
energies, this happens when $2\pi/\Delta_\mathrm{CMB}\lesssim L_\mathrm{c}$, or
\begin{equation}
  E\gtrsim 8\times 10^{12}\,{\rm eV}
  \left(\frac{L_\mathrm{c}}{10\unit{Mpc}}\right)^{-1}.
  \label{eq:E_coh}
\end{equation}
The coherence length of the intergalactic magnetic field is practically
unconstrained from above, for concreteness we will use
$L_\mathrm{c}\sim 5$\,Mpc as default value. Meanwhile, the Galactic magnetic
field has a turbulent component which coherence length is usually assumed to
be around $20$\,pc, and
the regular component should be comparable to the size of the Galaxy,
$\sim 10\unit{kpc}$. Thus, the
turbulent component of the Galactic magnetic field is expected to contribute
little to the ALP wiggles at CTA energies.

The ALP wiggles are most prominent around the transition from the strong mixing
regime, occurring when $  \Delta_\mathrm{CMB}\sim\Delta_{a\gamma}$ or
\begin{equation}
   E^\mathrm{crit}\simeq 2\times 10^{11}
    \unit{eV}\times\frac{g_{a\gamma}B}{10^{-11}\unit{GeV^{-1}}\unit{nG}}.
  \label{eq:ecrit}
  \end{equation}
Since CTA is most sensitive in the range between $10^{11}$ and $10^{14}$\,eV, 
one should ideally have $10^{11}\unit{eV}\lesssim E^\mathrm{crit}$. This yields
\begin{equation}
  \frac{g_{a\gamma}B_\mathrm{T}}{10^{-11}\unit{GeV}\unit{nG}}
  \lesssim 1/2.
  \label{eq:gB}
\end{equation}
Furthermore, the ALP mass should be small enough that there exists a strong
mixing
regime for the given magnetic field strength. This leads to the
condition $m_a\lesssim 10^{-10}B_\mathrm{T}/\mathrm{nG}$.
Since the onset of the wiggles is determined by the weakest magnetic field
and photon spectra are usually steeply falling, the
intergalactic magnetic field may prove to lead to the strongest wiggles.

In Eqs.~\eqref{eq:ecrit} and \eqref{eq:gB}, only the combination
$g_{a\gamma}B_\mathrm{T}$ is of importance. This implies that if we change
$g_{a\gamma}$ or $B_\mathrm{rms}$, we will change the energy at which the
wiggles are most prominent---weaker magnetic fields or lower $g_{a\gamma}$
implies that we should look at lower energies with a different detector.
However, the exact morphology and distribution of magnetic field strengths
is unknown. In general, a combination of the
Galactic magnetic field ($B=\mathcal{O}(\mu\rm G)$), magnetic
fields around galaxy clusters ($B=\mathcal{O}(\mu\rm G)$), and the
extragalactic magnetic field
($B=10^{-9}\text{--}10^{-14}\unit{G}$) will influence ALP oscillations.
Therefore, in the next two sections, we introduce and advocate for a search for
ALP wiggles in which the magnetic field does not need to be specified.
The example parameters that will be used
are chosen such that the wiggles will be prominent in the CTA energy range
($g_{a\gamma}=10^{-11}\unit{GeV}$,
$B_\mathrm{rms}=10^{-9}$ [Sec.~\ref{sec:stat}],
$B_\mathrm{rms}=5\times 10^{-9}$ [Sec.~\ref{sec:detection}]), and the results
should therefore only be considered as a proof of principle.

\section{Statistical tests for ALP wiggles}
\label{sec:stat}

\subsection{The $\chi^2$ test for irregularities}

The ALP wiggles induced by photon-ALP oscillations will be perceived as
``irregularities'' in the photon spectrum. Thus, one can use as a probe
the $\chi^2$ test,
\begin{equation}
  \chi^2 = \frac{1}{N_\mathrm{bins} - 1}
  \frac{[f_\mathrm{data}(E) - f(E)]^2}{\sigma_\mathrm{data}^2},
  \label{eq:TS_spec}
\end{equation}
where $f_\mathrm{data}(E)$ is the measured binned energy spectrum
(with photon-ALP oscillations if they exist), $\sigma_\mathrm{data}$ is
the experimental uncertainty, $N_\mathrm{bins}$ is the number of data points,
and $f(E)$ is the modelled spectrum
(without ALP oscillations)~\cite{Wouters:2012qd}.
However, even though this method is statistically sound, it can only measure
whether the photon spectrum is more irregular than statistically expected.
In the simulated examples in this work we `model' instead the spectrum by fitting
the function
\begin{equation}
  f(E)\propto E^{-b}\exp\left\{-\tau(E)\right\},
  \quad \text{ with } \quad \tau=\exp\{\beta(\log(E))\}
  \label{eq:fit_func}
\end{equation}
where $\beta(x)$ is a fifth order polynomial
and $b$ is the spectral index\footnote{Due to the high degeneracy of the fit,
we fix for simplicity $b$ to the simulated value. }
to the un-binned spectra
by minimising the maximum likelihood estimate (MLE) 
in order to isolate the effect of the wiggles in a model independent way.

\subsection{The discrete power spectrum}

The photon-ALP oscillations will perturb the photon spectrum by energy
dependent oscillations, $k\sim \Delta_\mathrm{osc}$, even for a turbulent
magnetic field. At energies above the strong mixing
regime, the ALPs with thus lead to wiggles with $k\sim E$ in the observed
photon spectra. Likewise, below the strong mixing regime, the wiggles
have the wavenumber $k\sim E^{-1}$. In Ref.~\cite{Kachelriess:2021rzc}, we
suggested therefore to use the windowed discrete power spectrum,
\begin{equation}
  G_N(k)=\left|\frac{1}{N}\sum_\mathrm{events}\me^{\iu\eta k}\right|,
  \label{eq:G}
\end{equation}
to extract information on the wiggles.
The sum in Eq.~\eqref{eq:G} goes over the $N$ detected photon events.
Only photons with energies between $E_\mathrm{min}$ and $E_\mathrm{max}$ are
included, and we use $\eta=E/E_\mathrm{min}$ to resemble the expected energy
dependence of the wiggles above the strong mixing regime.
A similar concept was introduced in Ref.~\cite{Conlon:2018iwn}.
Importantly, one can use the discrete power spectrum to
search for ALPs without specifying the magnetic field.
However, for a turbulent magnetic field, the ALP
signal is a broadened peak whose location and width is a priori unknown.
While this makes a detection more challenging, it enables the extraction of
information on the magnetic field.

Note that the signal strength depend on the choice of $E_\mathrm{min}$: It
should be chosen close to the transition from the strong mixing regime, which
a priori is unknown. This means, on the other hand, that the
conditions~\eqref{eq:E_coh} and \eqref{eq:gB} can in principle be used to
deduce the ALP parameters from a detected photon-ALP oscillation signal:
The combination $g_{a\gamma}B_\perp$
will for most astrophysical environments determine the onset of the oscillations
$\Delta_\mathrm{CMB}=2\Delta_{a\|}$ (see Fig.~3 in
Ref.~\cite{Kachelriess:2021rzc}).
This means that $g_{a\gamma}B_\perp$
can be fixed by finding the value of $E_\mathrm{min}$ that optimises the
observed oscillations. The mass $m_a$ can likewise be determined by X-ray
measurements.

We consider the test statistic (TS) given by the goodness-of-fit measure
compared to an estimated background,
\begin{equation}
  \mathrm{TS} = \frac{1}{\Delta k}\int_0^{\Delta k}
  \frac{[G_N(k) - G_N^B(k)]^2}{\sigma_N^B(k)^2} \dd{k}.
  \label{eq:TS_G}
\end{equation}
Here, $G_N^B(k)$ and $\sigma_N^B(k)$ are the estimated background power
spectrum and its 1$\sigma$ variation (see Sec.~\ref{sec:method}).
We choose $\Delta k=6$ to reduce the contributions from random fluctuations
at large $k$. While Eq.~\eqref{eq:TS_G} shares similarities with the $\chi^2$
statistics, one should emphasise that one expects a longer tail in this test
statistics since we are integrating over a range in which there statistically
is expected to be random peaks, i.e. the probability that there is a random peak
at any $k$ is larger than the probability that there is a peak at a fixed $k$.
The TS can in principle be improved if the shape, position and width is
taken into account, using for example machine learning.

\subsection{Statistical procedure and examples}
\label{sec:method}

In this section, the use of the discrete power spectrum to detect ALP wiggles
will be exemplified. We will focus on the effect of
the magnetic field modelling on the ALP wiggles, thereby illustrating the
importance of a proper treatment of the magnetic fields
in modelling photon-ALP oscillations.
Based on the discussions above, we define the following statistical procedure:
\begin{enumerate}
  \item Photons are sampled according to the chosen source spectrum
    using {\tt ELMAG}. The simulations are stopped when a given number $N$ of
    photons has reached the detector within the considered energy range.
    The energy of the simulated photons that reached the detector is used
    to compute the discrete power spectrum $G$.
  \item The ``measured spectrum'' is modelled by minimising the
    maximum-likelihood-estimate (MLE) of the fit function~\eqref{eq:fit_func}
    to the simulated data.
  \item The background power spectrum and its statistical variation is in turn
    found by drawing $N\times 10^3$ energies using the fitted
    spectrum as a probability distribution. 
  \item The TS is computed using \eqref{eq:TS_G}.
\end{enumerate}
In all the scenarios considered in this section, we repeat this procedure for
$10^3$ realisations of the magnetic field in order to obtain the distribution
of TS values.

\begin{figure}[htb]
  \centering
  \includegraphics{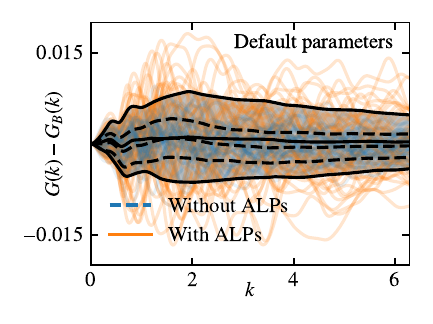}
  \includegraphics{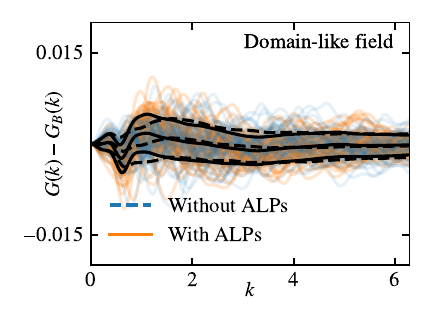}
  \caption{
    The power spectrum with the estimated background subtracted is plotted
    using a Gaussian turbulent field (left) and a domain-like field (right).
    The results for 50 realisation of the magnetic field with (orange) and
    without (blue) photon-ALP oscillations is shown, and the averages and the
    statistical standard deviations from a sample of $10^3$ realisations are
    shown in black lines.
    The parameters used in the simulations are discussed in the main text.
  }
  \label{fig:G}
\end{figure}

For concreteness, we will consider $N=10^4$ detected photons and in the energy
range $E\in( 10^{12}, 10^{14})\unit{eV}$ with the injection spectrum
$\dv*{N}{E}\propto E^{-1.2}$. 
Moreover, we fix $g_{a\gamma}=10^{-11}\unit{GeV^{-1}}$ and use a turbulent
magnetic field with $B_\mathrm{rms}=5\unit{nG}$. The power spectrum using
a Gaussian turbulent field with $\gamma = 5/3$, $\beta\to \infty$ and
$L_\mathrm{c}=5$\,Mpc (default parameters) is shown in the left pane of
Fig.~\ref{fig:G}. In all plots in this section, the various scenarios will be
labelled using the parameters that differs from the default parameters.
The results for 50~realisations with photon-ALP oscillations are shown in 
orange lines, those without photon-ALP oscillations in blue. The averages and
the $1\sigma$ statistical variance (black lines) were computed using the full
set of $10^3$ realisations. For comparison, the results using a domain-like
field are shown in the right pane of Fig.~\ref{fig:G}.

\begin{figure}[htb]
  \centering
  \includegraphics{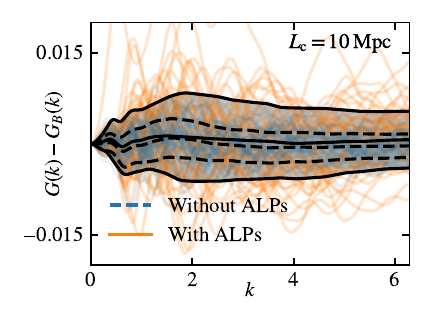}
  \includegraphics{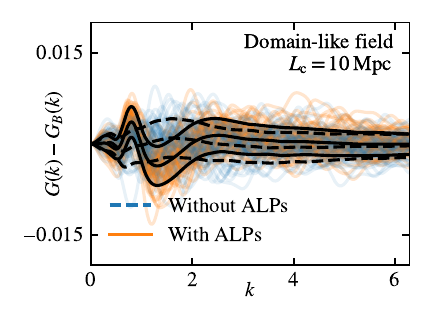}
  \caption{Same as Fig.~\ref{fig:G}, but with $L_\mathrm{c}=10$\,Mpc.}
  \label{fig:G10}
\end{figure}

The results in Fig.~\ref{fig:G} show the power of the statistical
procedure: There are clear peaks in the power spectrum including photon-ALP
oscillations compared to the case without photon-ALP oscillations%
\footnote{
  Note that there is a minor bump, still comparable with flat, in the power
  spectrum without any photon-ALP oscillations. This indicates that our fit
  function does not perfectly describe the optical depth of the used EBL model. 
   For the purposes of this paper,
  where the fitting procedure is made automatic using
  $\mathcal{O}(10^4)$ spectra, the quality of the fit is sufficient.
}.
Interestingly, due to the lack of cosmic variance in the simple domain-like
field, there is a lack of variance in the photon spectra which represents
itself as a clear signal in the discrete power spectrum, even after averaging
over many realisations of the magnetic field. 
This becomes even clearer for
larger coherence lengths, as shown in Fig.~\ref{fig:G10}
for $L_\mathrm{c}=10$\,Mpc. As such, the use of simplified magnetic field
models,
such as the domain-like field, may lead to a bias in searches for ALP wiggles
and impact the estimated limits on $g_{a\gamma}$. 
However, the larger variance in more realistic magnetic field
models---in these examples represented by Gaussian turbulent fields---increases
the rate of random encounters of regions of magnetic fields that may enhance
the wiggles (see also the discussion in, e.g., Ref.~\cite{Carenza:2022zmq}).
Thus, a more realistic
modelling of the magnetic fields may, in fact, improve the detection
prospects by such random encounters.
The detection prospects could be further improved choosing more suitable
fitting functions. Moreover, a constant windowing function was used. By
varying the minimal energy, $E_\mathrm{min}$, one may hope to further
increase the detection sensitivity

\begin{figure}[htb]\centering
  \includegraphics{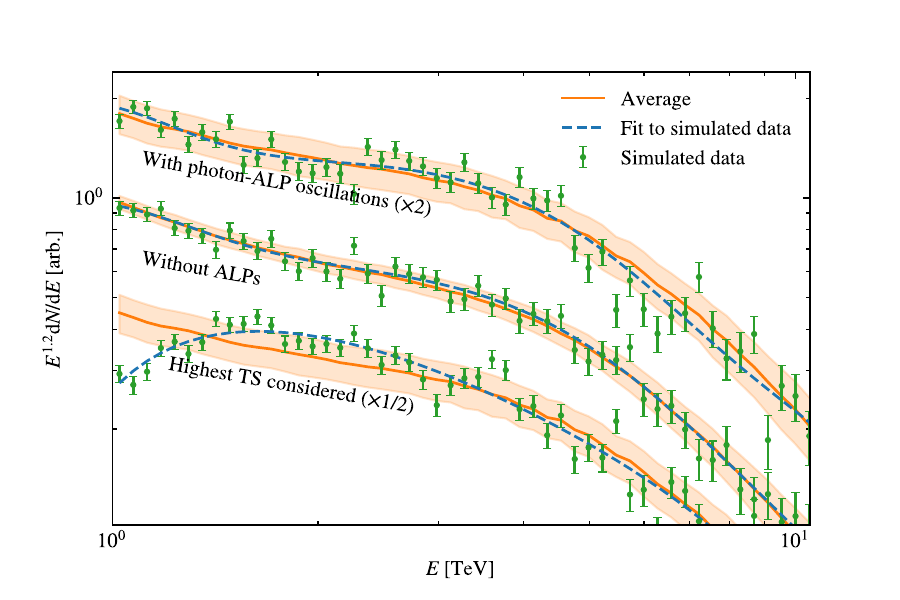}
  \caption{
  The simulated data (green errorbars) for one random realisation of the
  magnetic field are plotted with and without photon-ALP oscillations.
  The spectra are multiplied by a constant to improve visibility. 
  Furthermore, the function~\eqref{eq:fit_func} fitted to the (un-binned) data
  is shown as a blue dashed line together with the average obtained from
  the complete sample of 1000 realisations is shown.
  In addition, the spectrum from the simulation that yielded the highest TS is
  shown.
  }
  \label{fig:spectrum}
\end{figure}

For visualisation and to better understand the essence of the method, we plot
in Fig.~\ref{fig:spectrum} the binned energy spectrum (green errorbars) and
the fitted\footnote{
  We emphasise that un-binned data are used in the fit.
}
spectrum (blue dashed line) for one random realisation of the magnetic field,
both with and without photon-ALP oscillations. In addition, the spectrum
averaged over all simulations and its standard deviation is shown (orange
region). It is clear that photon-ALP oscillations increase the variation in the
energy spectrum.
The task of the generic fitting function~\eqref{eq:fit_func} is to reduce the
effect of unknown features in the source spectrum, such as
uncertainties in the modelling of the EBL or unresolved features in the 
source spectrum. This leads to a caveat of this approach, well visualised in
Fig.~\ref{fig:spectrum} with the spectrum that yielded the highest TS value in
this analysis: The spectrum may be ``over-fitted'', i.e.\ part of the signal
will be incorporated into the fit function, weakening thereby  the signal.
This applies especially for the wiggles extending over a larger energy range.
Since the true injection spectrum of the source is not known, a detailed
modelling of the source would be required in such cases to distinguish
between  intrinsic and ALP induced features in the  energy spectrum.

\begin{figure}[htb]\centering
  \includegraphics{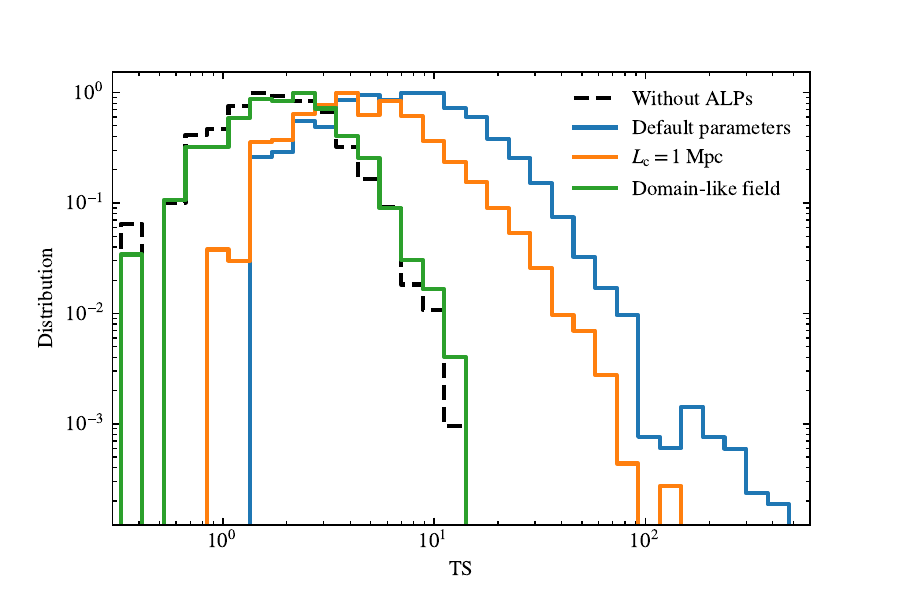}
  \caption{
    Histograms of the TS~\eqref{eq:TS_G} obtained using the statistical method
    described in subsection~\ref{sec:method}.
    The various colored lines are obtained using different parameters
    for the magnetic field; the labels indicate the parameter changed
    compared to the default parameters (see the main text for a description).
    The results in the no-ALP scenario is plotted as a dashed black line.
  }
  \label{fig:TS}
\end{figure}

In Fig.~\ref{fig:TS}, we plot the distribution of the TS~\eqref{eq:TS_G}, for
the default parameters, $L_\mathrm{c} = 1$\,Mpc,
$L_\mathrm{c}=10$\,Mpc, and a domain-like field.
With the chosen TS, the domain-like field is difficult to distinguish from the
non-ALP scenario. The Gaussian turbulent field, however, has a clear tail in the
TS distribution, which distinguishes the ALP from the non-ALP scenario. 
While increasing the coherence length improves the detection prospects,
details of the magnetic field like the values of $\gamma$ and $\beta$ have
only a minor influence on the TS distribution and we therefore do not
vary them in the figure. The reason for the weak dependence on these
parameters is that the integrated magnetic field distributions,
or the filling factor, is independent of the magnetic field spectrum for a
Gaussian turbulent field with the same $B_{\rm rms}$ and  $L_{\rm c}$.
In order to more clearly quantify the
differences, we list in Tab.~\ref{tab:C95} the probability that a signal is
detected with a confidence level of $2\sigma$, denoted as $C_\mathrm{95}$, and
the 99 \% quantile for
the various magnetic field scenarios considered. 
Although there are only minor differences in the $C_{95}$ value, there are
noticeable differences in the tails of the distributions, registered
in the 99\,\% quantiles. 

{\small 
\begin{table}[htb]
  \centering
\begin{tabular}{ccc}
  \hline
  Parameter          & $C_{95}$ & 99 \% quantile\\
  \hline
  Default            & 0.984  & 98.3 \\
  $\gamma=4/3$       & 0.989  & 135 \\
  $\gamma=2$         & 0.989  & 101 \\
  $\beta=2$          & 0.988  & 183 \\
  $\beta=4$          & 0.991  & 183 \\
  $f=0.1$            & 0.988  & 69.9 \\
  $L_\mathrm{c}=1\unit{Mpc}$  & 0.955 & 50.2 \\
  $L_\mathrm{c}=10\unit{Mpc}$ & 0.972 & 140 \\
  Domain-like field  & 0.628 & 9\\
  {\small \begin{tabular}{cc}
  Domain-like\\$L_\mathrm{c}=10\unit{Mpc}$\end{tabular}} & 0.660 & 10 \\
  \hline
\end{tabular}
  \caption{The probability that a signal is
    detected with a confidence level of $2\sigma$, denoted as $C_\mathrm{95}$,
    for the various magnetic field scenarios considered with Eq.~\eqref{eq:TS_G}
    as TS. 
  }
\label{tab:C95}
\end{table}
}

Note that the results of our examples clearly show that
the sensitivities depend strongly
on whether a Gaussian turbulent or a domain-like field is used:
Due to the cosmic variance in more realistic magnetic field models,
there is a chance that photons and ALPs
propagate through a region of magnetic fields favourable for photon-ALP
oscillations, thus enhancing the probability for detection.
The same conclusion can be drawn by, e.g., the results in
Ref.~\cite{Montanino:2017ara} wherein limits where set using a domain-like
magnetic field and using cosmic MHD simulations.
Likewise, in Ref.~\cite{Carenza:2022zmq}, it was found that cosmic
MHD simulations have a larger change of such ``rare encounters'' than
a Gaussian turbulent field.
Thus, using a more realistic field than the Gaussian magnetic field
considered here may improve the sensitivities even further.

%

\section{Detecting ALP wiggles from \PKS\ with CTA}
\label{sec:detection}

\begin{figure}
  \centering
  \includegraphics[]{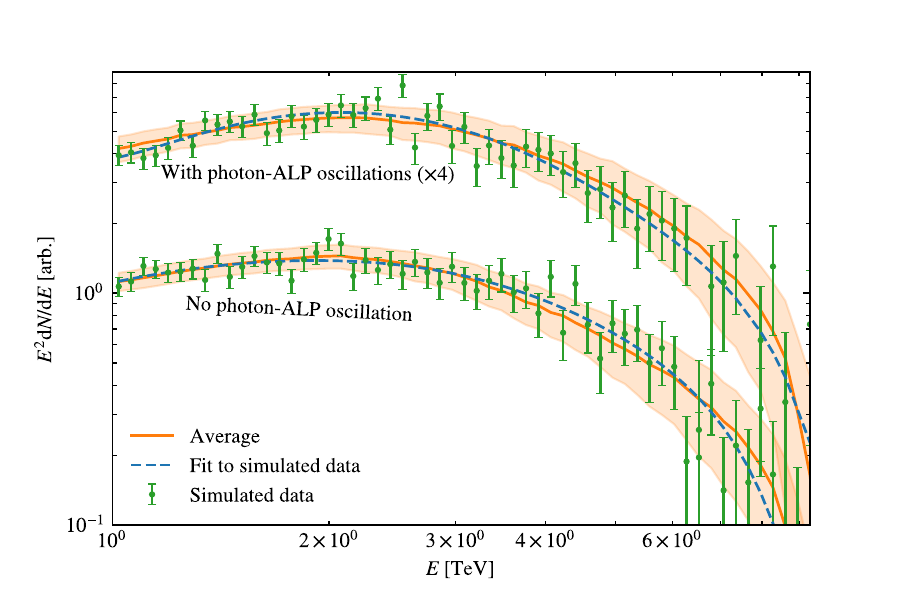}
  \caption{Same as Fig.~\ref{fig:spectrum}, but with $N=2.6\times 10^3$ photons
  and the parameters used for \PKS.}
  \label{fig:spec_PKS}
\end{figure}

In this section, we will consider \PKS~\cite{Aharonian:2007ig} at
redshift $z=0.116$ as a concrete example. Its photon spectrum can be
approximated by~\cite{Yu:2022psh}
\begin{equation}
  \dv{N}{E} = N_0\left(\frac{E}{E_b}\right)^{-\alpha - \beta\log(E/E_b)},
\end{equation}
with $N_0 = 15.4\times 10^{-12}\unit{cm^{-2}s^{-1}MeV^{-1}}$,
$E_b = 1136$\,MeV, $\alpha = 1.77$ and $\beta=0.035$. Since
CTA~\cite{Maier:2019afm} is not yet operational and its sensitivities are
preliminary, we assume conservatively an energy-independent effective collection
area $A=10^5\unit{m^2}$. We take, however, into account the energy resolution of
the detector by scrambling the detected energies using a normal deviate with an
energy dependent half-width given by the preliminary energy resolution of CTA.
Furthermore, we consider an energy range $E\in[10^{12},10^{14}]\unit{eV}$, for
which the energy resolution $\Delta E/E$ is approximately energy
independent\footnote{If
the larger energy range $E\in[10^{11},10^{14}]$\,eV is used, one
should convolve a parametrisation of the energy resolution with the fit
function.}.
The expected number of photons detected by CTA in this energy range
from \PKS\ can then be approximated as
\begin{equation}
  N = A\Delta t\int \dd{E} \dv{N}{E}
  \approx 2.6\times 10^3\factor{\Delta t}{50}{h} \factor{A}{10^5}{m^2},
\end{equation}
$A$ being the effective detector area and $\Delta t$ the detection
time.
In Fig.~\ref{fig:spec_PKS}, we plot the spectrum obtained from one simulation
of \PKS\ for an observation time $\Delta t=50$\,h
and a Gaussian turbulent field with $L_\mathrm{c}=5$\,Mpc.

\begin{figure}
  \centering
  \includegraphics[]{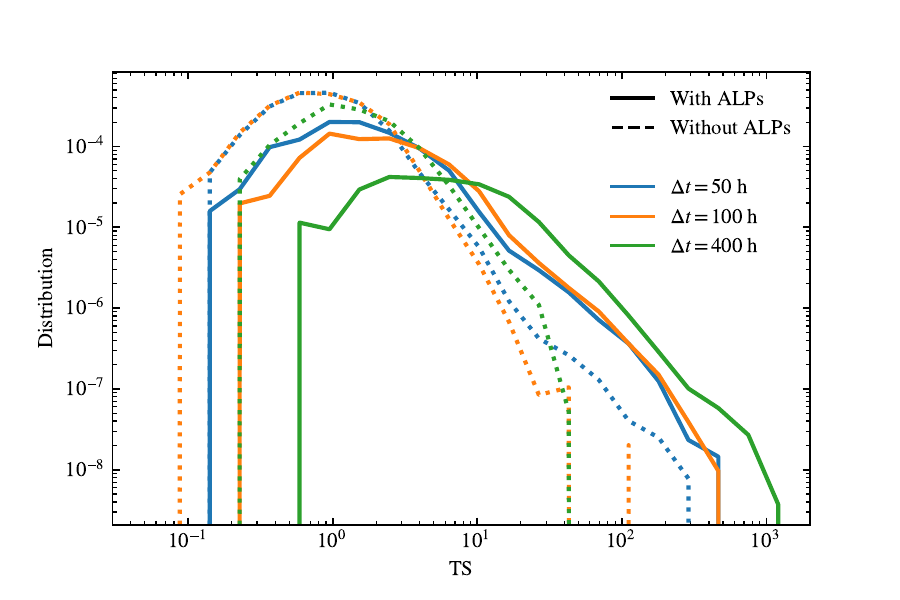}
  \caption{
    Histograms of the TS~\eqref{eq:TS_G} obtained using the statistical
    method described in subsection~\ref{sec:method} on the simulated data
    from {\PKS} for the observation times
    $\Delta t= 50$\,h (blue),
    100\,h (orange),
    200\,h (green) and
    400\,h (red).
    The corresponding no-ALP cases are shown with dashed lines.
  }
  \label{fig:TS_PKS}
\end{figure}

\begin{figure}
  \centering
  \includegraphics[]{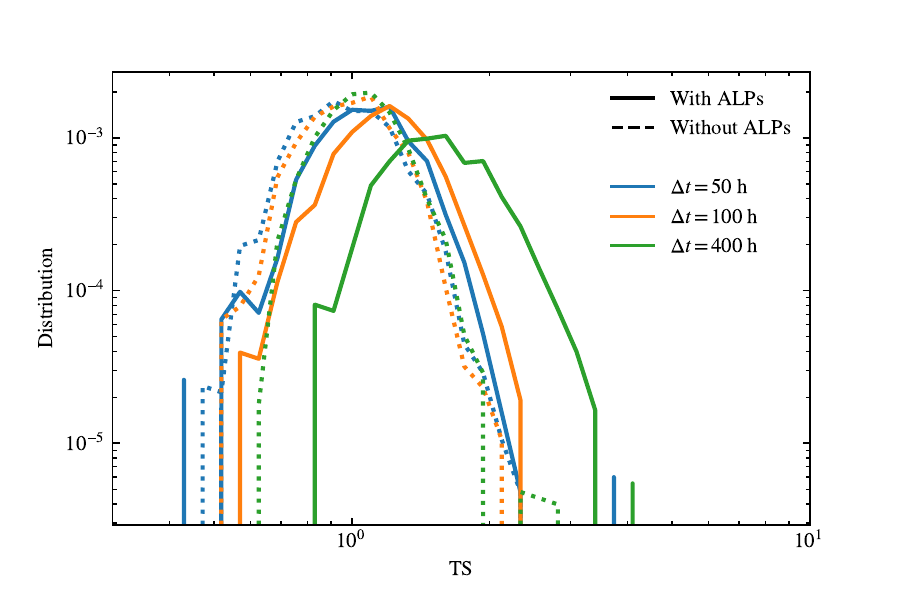}
  \caption{Same as Fig.~\ref{fig:TS_PKS}, but with the $\chi^2$ value
  [Eq.~\eqref{eq:TS_spec}] as TS.
  }
  \label{fig:TS_spec_PKS}
\end{figure}

To get an idea of the detectability of ALP wiggles from \PKS,
we follow the statistical procedure from section~\ref{sec:method}. The
result is shown in Fig.~\ref{fig:TS_PKS} for observation times
$\Delta t = \{50, 100, 400\}$\,h. As expected, 
increasing the observation time increases the detection prospects. 
As a basis for comparison, we consider in Fig.~\ref{fig:TS_spec_PKS} the TS
distribution using Eq.~\eqref{eq:TS_spec} with the same binning as in
Fig.~\ref{fig:spec_PKS}.


In order to more clearly quantify the
differences between the statistical methods used in Figs.~\ref{fig:TS_PKS}
and~\ref{fig:TS_spec_PKS},
we list in Tab.~\ref{tab:C95_PKS} the probability that a signal is
detected with a confidence level of $2\sigma$, denoted as $C_\mathrm{95}$, for
the various magnetic field scenarios considered. 
From these values, we conclude that
the use of the discrete power spectrum leads to better detection prospects
compared to a standard irregularity search in the energy spectrum,
especially for low observation times.
Note, however, that the search for irregularities will depend on the binning,
and can thus be improved. On the flip side, the optimal
windowing function for the power spectrum will depend on the ALP coupling and
the magnetic field strength. There is, in any case, a clear advantage of using
the discrete power spectrum compared to a standard search for residual: While a
high $\chi^2$ value merely indicates that the data are more
irregular than expected, a signal in
the discrete power spectrum is a clear indication that the data have
wiggles with
the same energy dependence as expected for photon-ALP oscillations.

{\small 
\begin{table}
  \centering
\begin{tabular}{c|cccc}
              & $\Delta t=50\unit{h}$ & $\Delta t = 100\unit{h}$ &
                $\Delta t = 200\unit{h}$ & $\Delta t = 400\unit{h}$ \\ \hline 
  Power spec. & 0.218                 & 0.469       & 0.591 & 0.644 \\
  Spec.       & 0.094                 & 0.228       & 0.478 & 0.597  \\
\end{tabular}
  \caption{The table indicates the probability that ALPs are detected in an
    observation of \PKS\ using CTA with a confidence
    level larger than $2\sigma$, denoted as $C_\mathrm{95}$, using the power
    spectrum [Eq.~\eqref{eq:TS_G}] and a standard $\chi^2$ search
    [Eq.~\eqref{eq:TS_spec}]. The columns corresponds to the detection
    times used in Fig.~\ref{fig:TS_PKS}.
  }
\label{tab:C95_PKS}
\end{table}
}

\begin{figure}
  \centering
  \includegraphics[]{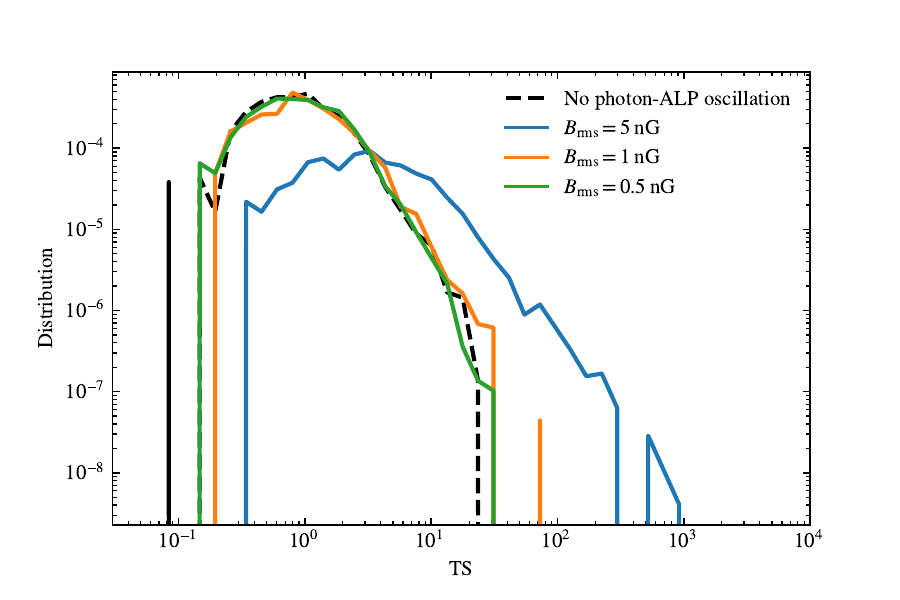}
  \caption{Histograms of the TS~\eqref{eq:TS_G} obtained using the statistical
    method described in subsection~\ref{sec:method} on the simulated data
    from {\PKS} for an observation time $\Delta t=200$\,h. 
    The results obtained with magnetic field strengths
    $B_\mathrm{rms}=5\unit{nG}$ (blue),
    $B_\mathrm{rms}=1\unit{nG}$ (orange) and
    $B_\mathrm{rms}=0.5\unit{nG}$ (green)
    are shown.
    The no-ALP scenario is shown in dashed line.}
  \label{fig:TS_PKS_B}
\end{figure}

In this section, we have considered an optimistic value for the
intergalactic magnetic field, $B_\mathrm{rms} = 5\times 10^{-9}$\,G, and 
an excluded coupling strength, $g_{a\gamma}=10^{-11}\unit{GeV^{-1}}$.
It is thus useful to check to what
extent the detectability worsens when the magnetic field strength is
decreased\footnote{
  Since the oscillation depend on the magnetic field and the
  coupling strength via the combination $g_{a\gamma}B$, this is equivalent to
  reducing the coupling.
}.
Therefore, we plot in Fig.~\ref{fig:TS_PKS_B} the histograms of the TS obtained
for varying magnetic field strength, $B_\mathrm{rms}=\{5, 1, 0.5\}\unit{nG}$.
The corresponding detection probabilities are $C_\mathrm{95}=\{0.947, 0.547,
0.521\}$.
Two effects lead to the quick reduction in $C_\mathrm{95}$ with decreasing 
magnetic field strength: First, the wiggles are strongest close to the strong
mixing regime, and decreasing the magnetic field strength shifts the strong
mixing regime to lower energies. From the condition in Eq.~\eqref{eq:gB}, it it
expected that our choice of default parameters leads to the strongest wiggles.
This reduction in sensitivity may be partly compensated for by changing the
lowest energy considered,
for example by considering a different detector.
Second, the mixing strength is proportional to the magnetic field strength,
which can only be compensated for by increasing the observation time.
At lower energies, however, the number of photons is usually significantly
higher.

\section{Summary and conclusion}

Photon-ALP oscillations will imprint energy-dependent oscillatory features,
which we name ``ALP wiggles'', on photon spectra from distant high-energy
sources. We have therefore proposed to use the
discrete power spectrum~\eqref{eq:G} to directly probe such wiggles
in experimental data. Such a search will be independent of the modelling of
magnetic fields and theoretical uncertainties in, e.g., the EBL.
This work serves as a first proof of principle, and there is room for
improvement: We only considered the simple
test statistic~\eqref{eq:TS_G} which only measures the residual of the
measured discrete power spectrum compared to the estimated background.
Furthermore, in order to stay as concrete as
possible, and since the onset of the ALP wiggles at TeV energies is determined
by the weakest magnetic field that contribute to the oscillations, we considered
only an intergalactic magnetic field. In a complete analysis, one should
furthermore consider photon-ALP oscillations in, e.g.,
the Milky Way and the host galaxy, and in the source itself.
As a second step, the discrete power
spectrum can be used to extract information about the magnetic field, more
specifically, it can be related to the two-point correlation function of the
magnetic field~\cite{Kachelriess:2021rzc,Marsh:2021ajy}.

We have compared two different treatments of the magnetic field: a Gaussian
turbulent field, and a simple and unphysical domain-like field. We found
that the increased cosmic variance of the Gaussian turbulent field may
significantly improve the detection prospects. Varying  the shape of its
power spectrum, we did not observe a strong dependence
on the resulting axion wiggles,  as long as the effective coherence length
remained constant.

As a concrete example, we considered the detection of ALP wiggles in the
energy spectrum from \PKS\ using conservative estimates of the
Cherenkov Telescope Array (CTA) sensitivity.
Our analysis indicates that ALP wiggles can be detected by CTA for optimistic
values of the magnetic field and photon-ALP coupling. Importantly, the method
is an improvement compared to a standard search for ``irregularities'' in 
photon spectra, which is currently used to set leading limits.
The statistical method can furthermore
be optimised choosing an appropriate windowing
function:
Since the extragalactic magnetic field strength currently is only
weakly constrained, one cannot know at which energy the first ALP wiggle
occur.
Moreover, the simple test statistic  considered does
not take into account the size, shape and location of the peak.

\section*{Acknowledgements}
We would like to thank M.~Unger for valuable discussions that helped to
reduce the computational time required for generating the Gaussian
turbulent fields. JT would like to thank M.~Meyer for interesting discussions
and hospitality at the University of Hamburg. 
This article is based upon work from the COST Action COSMIC WISPers CA21106,
supported by COST (European Cooperation in Science and Technology).


\providecommand{\href}[2]{#2}\begingroup\raggedright\endgroup

\end{document}